\documentstyle[epsfig]{aipproc} 
 
\begin{document} 
\title{Etching of random solids: hardening dynamics and self-organized 
fractality} 
 
\author{A. Gabrielli$^{1,2}$, A. Baldassarri$^{3}$, and B. Sapoval$^{1,4}$} 
\address{$^{1}$Laboratoire de la Physique de la Matiere Condens\'{e}e,\\ Ecole
Polytechnique - CNRS, 91128 - Palaiseau, France\\
$^{2}$INFM - Dip. di Fisica, Univ. di Roma ``La Sapienza'',
P.le A. Moro, 2, ls I-00185 Roma, Italy\\
$^3$INFM - Dip. di Matematica e Fisica, Univ. di Camerino,\\ Via
Madonna delle Carceri, I-62032 Camerino, Italy\\
$^{4}$Centre de Math\'{e}matiques et de leurs Applications,\\
Ecole Normale Sup\'{e}rieure - CNRS, 94140 Cachan, France
} 
 
\maketitle

\begin{abstract} 
When a finite volume of an etching solution comes in contact 
with a disordered solid, a complex dynamics of the solid-solution 
interface develops. Since only the weak parts
are corroded, the solid surface hardens progressively.
If the etchant is consumed in the chemical reaction, 
the corrosion dynamics slows down and stops spontaneously leaving 
a fractal solid surface, which
reveals the latent percolation criticality hidden in any random system.
Here we introduce and study, both analytically and numerically,
a simple model for this phenomenon. 
In this way we obtain a detailed description of the process in terms of 
percolation theory. In particular we explain the mechanism of hardening of 
the surface and connect it to Gradient Percolation. 
\end{abstract} 
 
\section*{Introduction} 

Corrosion of solids is an everyday phenomenon with
evident practical consequences and applications \cite{uhlig}.
Recent developments of theoretical tools
for the study of disordered systems and fractals
in the context of statistical mechanics, has triggered
and outburst of activity in this subject.

When an etching solution is put in contact with a
disordered etchable solid, the solution corrodes
the weak parts of the solid surface while the
hard parts remain un-corroded. During this process 
new regions of the solid (both hard and weak)
come into contact with the etching solution.
If the volume of the solution is finite and the etchant is consumed
in the chemical reaction,
then the etching power of the solution diminishes progressively
and the corrosion rate slows down. 
When the solution is too weak to coorrode any part of the hardened
solid surface, the dynamics spontaneously stops. 
Note that the etchant concentration at the arrest time is not zero,
and, as shown below, its final value is strictly related to the 
percolation critical point.
One of the most interesting aspects of this
phenomenon is that the final connected solid
surface has a fractal geometry 
up to a certain characteristic scale $\sigma$ given by its thickness.
This is precisely the qualitative phenomenology that has been 
recently observed in a nice experiment on pit corrosion of
aluminum thin films \cite{Balazs}.
As we show below this fractality falls in the universality class
of percolation theory \cite{Aharony}.

\section*{The model}
Recently, a simple dynamical model of etching,
capturing the aforementioned phenomenology, has
been proposed and studied \cite{model,GBS}.
\begin{figure}[b!] 
\centerline{\epsfig{file=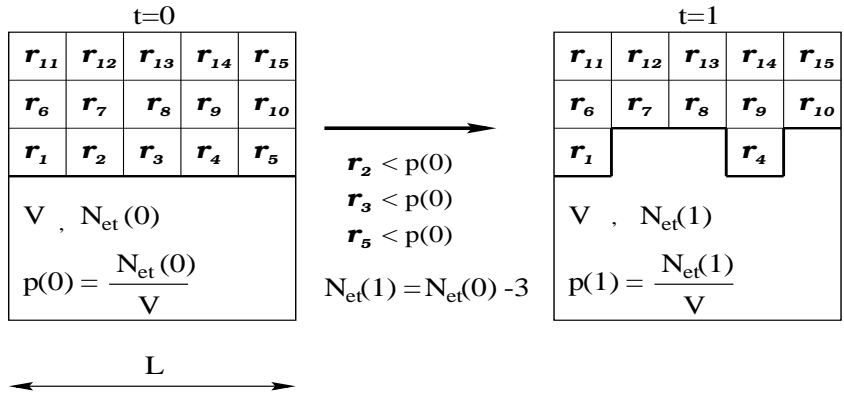, width=12cm, height=5.5cm}} 
\vspace{10pt} 
\caption{Sketch of the etching dynamics in a square lattice:
 the sites $2,3,5$ are etched
at the first time-step as their resistances are lower than $p(0)$.
At the same time the number of etchant particles 
in the solution decreases by $3$
units, and a new part of the solid is uncovered.} 
\label{fig1} 
\end{figure} 
The model is reviewed in Fig.~\ref{fig1} and described as follows:\\
(1) The solid is represented by a lattice of sites exhibiting random
``resistances to corrosion'' $r_{i}\in [0,1]$ uniformly distributed. 
It has a width $L$ and a given fixed depth $Y$ (in the 
case infinite). 
The etching solution has a finite constant volume $V$ ($V\gg L$) 
and contains an initial number 
$N_{et}(0)$ of etchant molecules. 
Therefore the initial etchant concentration is given by $C(0)=N_{et}(0)/V$.
At any time-step $t$ the ``etching
power'' $p(t)$ of the solution is proportional to the etchant
concentration $C(t)$. 
Hereafter we take the proportionality constant equal to one 
without loss of generality, and 
we choose $p_c<p(0)\le 1$, where $p_c$ is the percolation threshold of the 
lattice.\\  
(2) The solution is initially in contact with the solid through the bottom
boundary $y=0$. 
At each time-step $t$, all surface sites with $r_{i}<p(t)$ are
dissolved, and a particle of etchant is consumed for each corroded site.
Hence, the concentration of the solution decreases with $t$. 
Moreover, depending on the connectivity criterion chosen for the lattice,
$m(t)$ new solid sites, previously in the bulk, come into contact 
with the solution.
Note that at the next time-step, only these last sites can be corroded 
by the solution, as the other surface sites have already
resisted to etching and $p(t)$ decreases with $t$.
Calling $n(t)$ the number of dissolved solid sites at time $t$, one can write
\begin{equation}
p(t+1)=p(t)-\frac{n(t)}{V}\,.
\label{p-n}
\end{equation}
The corrosion dynamics spontaneously stops at the time-step $t_f$,
when all the surface sites have a resistance $r>p(t_f)$. 

The main characteristics of the model are:\\
(i) The final value $p_f=p(t_f)$ is slightly smaller than 
the percolation threshold $p_c$.
The difference $|p_f-p_c|\rightarrow 0$ as
$(L/V)^{\alpha_p}$ with $\alpha_p\simeq 0.45$,
when the limits $L\rightarrow +\infty$ and 
$V/L\rightarrow +\infty$ are taken in the right way \cite{GBS}.
This fact can be explained in terms of percolation theory. 
Indeed, this theory implies that,
for $p(t)\le p_c$, the probability to have a connected path 
of solid sites each one with $r>p(t)$, then stopping corrosion,
is equal to one in the right thermodynamic limit.
\begin{figure}[b!] 
\centerline{\epsfig{file=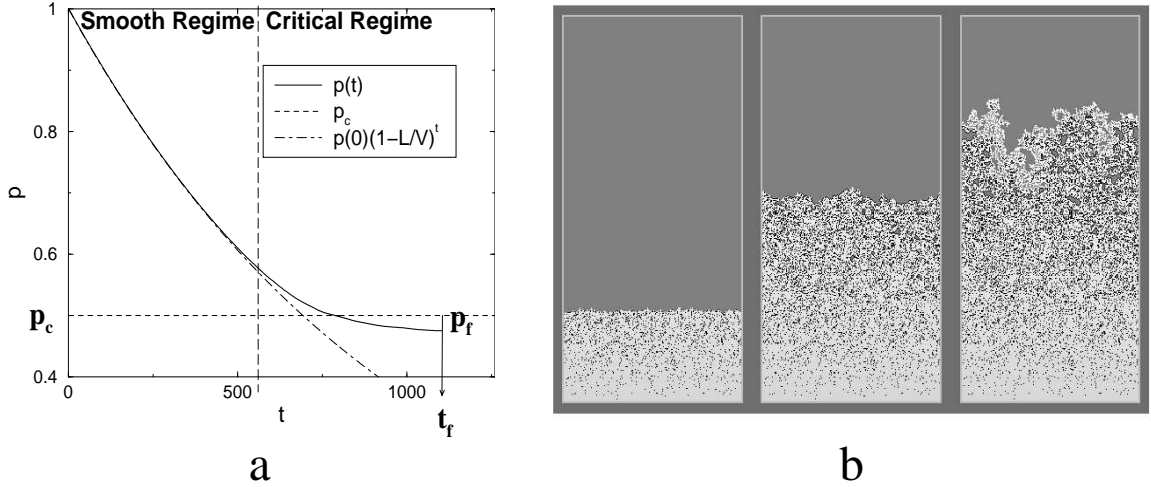, width=16cm}} 
\vspace{10pt} 
\caption{{\bf a}- Time decay of $p(t)$ (continuous line),
compared to the analytical approximation $p(t)=p_0\exp (-tL/V)$
in the smooth regime (dashed line). It refers to simulations with $p(0)=1$
and $V/L=1000$ in a triangular lattice. 
The two lines are in good agreement for $p(t)>p_c$. 
{\bf b}- Three snapshots of the corroded solid (dark grey) by the solution 
(clear) at three different time-steps: initial, intermediate and final.
Black regions represent detached islands. 
The final corrosion front shows fractality up to a characteristic length given
by its thickness.} 
\label{fig2} 
\end{figure} \\
 (ii) The global temporal behavior of $p(t)$ can be divided 
(see Fig.~\ref{fig2}-{\bf a})
into two regimes: a {\em smooth} regime when $p(t)>p_c$, and a {\em critical} 
regime when $p(t)\simeq p_c$. 
The duration of the smooth regime is measured approximately by $t_c$,
defined by $p(t_c)=p_c$. The quantity $t_c$ is found to scale
with the ratio $L/V$ in the following way $t_c\sim V/L$. 
The duration of the critical regime is given by $(t_f-t_c)$,
and scales as $(t_f-t_c)\sim (V/L)^{\alpha_{t_f}}$ with $\alpha_{t_f}\simeq 0.55$
(apart from a further linear dependence on $\log\,L$ of the scaling coefficient
due to the extremal nature of $t_f$ \cite{frac}).\\
(iii) The global surface of the solid in contact with the solution displays a 
peculiar roughening dynamics (three snapshots at almost initial, intermediate,
and final time are reported in Fig.~\ref{fig2}-{\bf b}).
First of all, during the process, the solution can surround and detach finite
solid islands. 
Consequently, the global solid surface is composed by the sum of
the perimeter of the ``infinite'' solid, here called {\em corrosion front}, and
the surfaces of the finite islands. 
In the first smooth regime, the corrosion is well directed and
the corrosion front becomes progressively rougher and rougher, while 
finite detached islands are quite small. 
In the second critical regime, the correlations revealed by the
hardening process become important: the dynamics becomes 
locally isotropic generating a fractal front. Detached islands
in this last period are quite big with a fractal perimeter too.
\begin{figure}[b!] 
\centerline{\epsfig{file=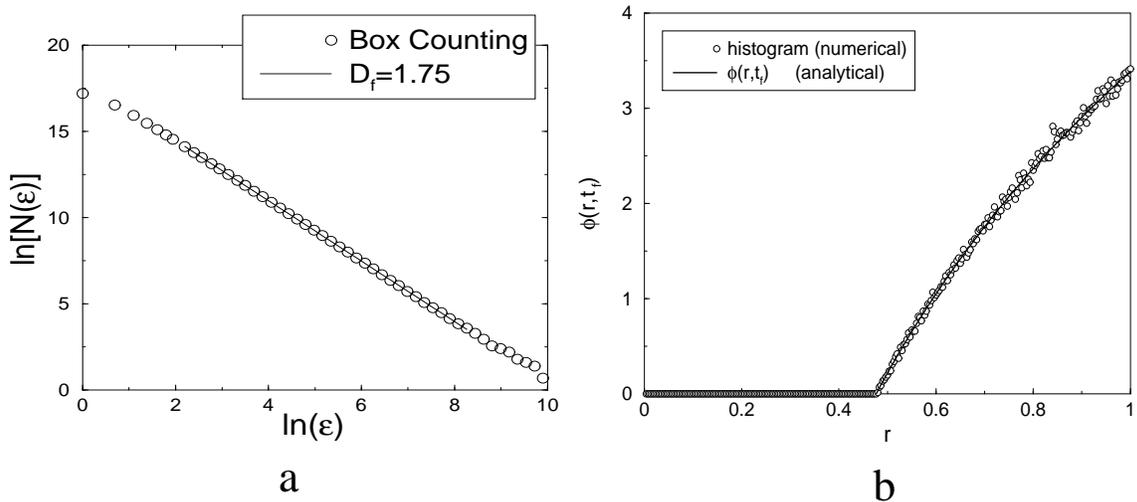, width=15cm}} 
\vspace{10pt} 
\caption{{\bf a}-Box counting measure of the fractal dimension of the corrosion 
front for length smaller than $\sigma$. {\bf b}-Final normalized histogram of
the global solid surface: simulations results (empty points) are compared
with the analytical evaluation (continuous line). 
} 
\label{fig3} 
\end{figure} \\
(iv) The final corrosion front is fractal with a fractal dimension
$D_f=1.753\pm 0.05$ (see Fig.~\ref{fig3}-{\bf a}) up to a characteristic
scale $\sigma$ which is the final thickness of the front. 
Beyond this length the corrosion front acquires a one-dimensional shape.
Also $\sigma$ is found to scale with the ratio $L/V$ in the following way:
$\sigma\sim (L/V)^{-\alpha_{\sigma}}$ with 
$\alpha_{\sigma}\simeq 0.57\simeq 1/D_f$.
As shown below, all these properties can be explained in the framework of 
percolation theory and in particular of Gradient Percolation \cite{GP}.\\
(v) Another important feature captured by this 
simple model is the progressive hardening of the solid surface, 
due to the corrosion only of sites with a resistance weaker than the etching
power of the solution, which, in its turn, decreases with time.
This hardening effect is described quantitatively
by the behavior of the normalized resistance histogram 
$\phi(r,t)$ of the global surface sites.
Obviously, $\phi(r,0)=1$ with $r\in [0,1]$, while $\phi(r,t_f)$ is given in 
Fig.~\ref{fig3}-{\bf b}. As one can see, the final global surface
is much more resistant to corrosion than the initial one.

\section*{Analytical result and percolation}
In order to have a mathematical insight in the previous numerical results, 
and a more clear
explanation in terms of percolation, it is useful to study the temporal evolution
of the above defined resistance histogram.
Let be $h(r,t)dr$ the number of surface sites with resistance in the interval 
$[r,r+dr]$ at time $t$ ($\phi(r,t)$ is the normalized version of $h(r,t)$).
Obviously, the number of corroded sites at time $t$ is 
$n(t)=\int_0^{p(t)}dr\,h(r,t)$.
Let us define $\omega(t)=m(t)/n(t)$. Then, using Eq.~(\ref{p-n}), 
we can write the two coupled equations:
\begin{eqnarray}
&&h(r,t+1)=h(r,t)-h(r,t)\theta[p(t)-r]+\omega(t)\int_0^{p(t)}dr'\,h(r',t)
\nonumber\\
&&p(t+1)= p(t)-\frac{\int_0^{p(t)}dr'\,h(r',t)}{V}\,,
\label{system}
\end{eqnarray}
where $\theta(x)$ is the usual step Heavyside function.
In the first relation of Eq.~(\ref{system}), the second term on the right side 
represents the number of just corroded sites with resistance between $r$ and 
$r+dr$, 
while the third one gives the contribution to
the new histogram coming from the $m(t)$ new surface sites.
All the geometrical features of the system related to the behavior of 
$p(t)$ and $h(r,t)$ are described by the coefficient $\omega(t)$.
In order to close this system of equations, we introduce two different 
approximations
of $\omega(t)$ for the smooth and the critical regimes.
One can show \cite{GBS} that, since in the smooth regime $n(t),m(t)\gg 1$ 
and fluctuations are small, the right approximation is
$\omega(t)=1/p(t)$. 
Using this relation in Eq.~(\ref{system}), one gets 
\begin{equation}
p(t)=p(0)\exp(-t/\tau)\,,
\label{p-lin}
\end{equation} 
with $\tau=-1/\ln (1-L/V)\simeq V/L$.
The good accuracy of this approximation in the smooth regime is shown
by Fig.~\ref{fig2}-{\bf a}.  
Eq.~(\ref{p-lin}) gives $t_c\sim V/L$ in good agreement with simulations.
In the critical regime, the only simple approximations is a mean-field-like
relation $\omega(t)=1/p_c$ \cite{GBS}, forgetting fluctuations which 
are important in this regime.
Putting this relation inside Eq.~(\ref{system}) one obtains, 
$(p_c-p_f)\sim \sqrt{L/V}$ and $(t_f-t_c)\sim \sqrt{V/L}$ which are not very
far from the simulation behaviors. Finally, the histogram $h(r,t)$
obtained by 
Eq.~(\ref{system}) within these approximations is in excellent agreement with 
simulations (see Fig.~\ref{fig3}-{\bf b}).
 
The link between this dynamical etching model and percolation criticality 
can be shown by analyzing more deeply Eq.~(\ref{p-lin}) valid in the smooth 
regime. 
From Eq.~(\ref{p-lin}), one has $t_c\simeq V/L\ln [p(0)/p_c]$. This implies that
the condition for smooth regime is $t<V/L$.
Then in this temporal range one can linearize Eq.~(\ref{p-lin}):
\[p(t)\simeq p(0)(1-t\,L/V)\,.\]
Since in this regime the solution advances mainly layer by layer 
into the solid, one can say that the layer etched at time $t$ is at a
depth $y=t$. 
We can map then our dynamical model in a static model defined by 
the following rule: 
a general solid site belonging to the layer $y$ is replaced by the solution
with a $y$-dependent probability $p(y)\simeq p(0)(1-y\,L/V)$.
This static model, if extended to any value of $p(y)>0$, is known under the name
of Gradient Percolation (GP) \cite{GP}.
In our case the gradient $\nabla p(y)$ of replacement (etching) 
probability satisfies $\nabla p\sim L/V$.
Why this mapping can be extended also out of the smooth regime is more clear
from field theory arguments \cite{Munoz} .
The GP model is characterized by the following features:
the boundary (the corrosion front in our case) 
between the connected etched region and the connected non-etched one 
has fractal features in the universality class of critical percolation.
In particular it has a fractal dimension $D_f=7/4$ coinciding
with the hull fractal dimension of infinite percolation cluster at 
criticality \cite{Duplantier}.
The fractality is not extended to the whole corrosion front, but
it is restricted to regions of size $\sigma\sim \nabla p^{-1/D_f}$, 
which, considering $\nabla p\sim L/V$, is in good agreement with our numerical 
results.
Furthermore, the range of $p$'s on this boundary is given by $\Delta p\sim
\sigma\nabla p\sim (L/V)^{1-1/D_f}$, which is well confirmed by simulation
considering the quantity $p_c-p_f$.
Note that from previous scaling relations,
one can write $\sigma\sim \Delta p^{-\nu}$, where $\nu=1/(D_f-1)=4/3$ is the
usual correlation length exponent in percolation. 
This shows that in the dynamical etching model $\sigma$ can be interpreted as the 
correlation length of the corrosion front: that is, the fact that $p_f$ is
not exactly equal to $p_c$ introduce a finite correlation length in the problem. 

This work has been supported by the European Community TMR Network
``Fractal Structures and Self-Organization" ERBFMRXCT980183.

\end{document}